\documentclass[journal,onecolumn]{IEEEtran}
\usepackage{graphicx}
\usepackage[dvipsnames]{xcolor} 
\usepackage{tabularx}
\usepackage[compress,nospace]{cite}
\usepackage{bchart}
\usepackage{booktabs}
\usepackage{multirow}
\usepackage{amssymb}
\usepackage{pifont}
\usepackage{array,colortbl}
\usepackage{adjustbox}
\usepackage{amsfonts}
\usepackage{multirow}
\usepackage{tikz}
\usepackage{float}
\usepackage{textcomp}
\usepackage{braket}
\usepackage{amsmath,calc}
\usepackage{subcaption}
\usepackage{hhline}
\usepackage{array, makecell}
\usepackage{boldline}
\usepackage{caption}
\usepackage{booktabs}
\usepackage[flushleft]{threeparttable}
\usepackage{amsmath}
\usepackage[utf8]{inputenc}
\usepackage{tikz}
\usetikzlibrary{shapes.geometric, arrows}
\usepackage{colortbl}
\usepackage{fixltx2e}
\usepackage{subcaption}
\usepackage{enumitem}
\usepackage{dblfloatfix}
\usepackage{authblk}
\usepackage[flushleft]{threeparttable}
\graphicspath{{./figures/}}

\DeclareCaptionLabelFormat{andtable}{#1˜#2 \& \tablename˜\thetable}

\newcolumntype{R}[2]{%
	>{\adjustbox{angle=#1,lap=\width-(#2)}\bgroup}%
	l%
	<{\egroup}%
}

\usepackage{textcomp}
\renewcommand{\thetable}{\arabic{table}}

\makeatletter
\def\endthebibliography{%
	\def\@noitemerr{\@latex@warning{Empty `thebibliography' environment}}%
	\endlist
}

\newcommand*{\rom}[1]{\expandafter\@slowromancap\romannumeral #1@}

\begin{document}

\title{\vspace{-0.5in}  Limitations of ROC on Imbalanced Data: Evaluation of LVAD Mortality Risk Scores}

\author[1]{Faezeh Movahedi}
\author[2]{Rema Padman}
\author[3]{James F Antaki}

\affil[1]{Swanson School of Engineering, University of Pittsburgh, Pittsburgh, PA}
\affil[2]{Heinz college, Carnegie Mellon University, Pittsburgh, PA}
\affil[3]{Meinig School of Biomedical Engineering, Cornell University, Ithaca, NY}
\maketitle

\begin{abstract}
\noindent
\textbf{Objective:} This study illustrates the ambiguity of ROC in evaluating two classifiers of 90-day LVAD mortality. This paper also introduces the precision recall curve (PRC) as a supplemental metric that is more representative of LVAD classifiers performance in predicting the minority class.\\

\noindent
\textbf{Background:} In the LVAD domain, the receiver operating characteristic (ROC) is a commonly applied metric of performance of classifiers. However, ROC can provide a distorted view of classifiers ability to predict short-term mortality due to the overwhelmingly greater proportion of patients who survive, i.e. imbalanced data.\\

\noindent
\textbf{Methods:} This study compared the ROC and PRC for the outcome of two classifiers for 90-day LVAD mortality for 800 patients (test group) recorded in INTERMACS who received a continuous-flow LVAD between 2006 and 2016 (mean age of 59 years; 146 females vs. 654 males) in which mortality rate is only $\%8$ at 90-day  (imbalanced data).  The two classifiers were HeartMate Risk Score (HMRS) and a Random Forest (RF).\\

\noindent
\textbf{Results:} The ROC indicates fairly good performance of RF and HRMS classifiers with Area Under Curves (AUC) of $0.77$ vs.  $0.63$, respectively. This is in contrast with their PRC with AUC of $0.43$ vs. $0.16$ for RF and HRMS, respectively. The PRC for HRMS showed the precision rapidly dropped to only $10\%$ with slightly increasing sensitivity.\\

\noindent
\textbf{Conclusion:} The ROC can portray an overly-optimistic performance of a classifier or risk score when applied to imbalanced data. The PRC provides better insight about the performance of a classifier by focusing on the minority class.\\

\end{abstract}

\section{Introduction}	 
Recent advances in predictive modeling have opened up new horizons for accurate therapeutic decision-making in clinical fields such as Mechanical Circulatory Support (MCS) device therapy for patients with advanced heart failure. 
Modern predictive models have the ability to discover the relationship between risk factors prior to receiving an LVAD and outcomes such as survival or freedom from adverse events, that are otherwise too complicated to be learned by a human. Examples of predictive models for LVAD outcomes include risk scores such as the HeartMate \rom{2} Risk Score (HMRS) \cite{cowger2013predicting}, Destination Therapy Risk Score (DTRS) \cite{lietz2007outcomes}, and the Penn-Columbia Risk Score \cite{birati2015predicting}. These commonly are based on a small collection of clinical variables. More recently predictive models have been introduced that consider hundreds of features such as Cardiac Outcome Risk Assessment (CORA) \cite{loghmanpour2015new}.

Nevertheless, no model is ever perfect, therefore it is essential to consider their limitations and predictive accuracy when used clinically. Accordingly, when introducing a new model, it is common, indeed best practice, to report various metrics of performance using, for example, a Cox model and ROC \cite{cowger2013predicting,birati2015predicting,ravichandran2015left,schaffer2009evaluation}. However, in the context of MCS, there is an important consideration that is often, if not always overlooked: namely the imbalanced distribution of outcomes. Due to the increasing success of LVAD therapy, 90-day survival has approached more than $90\%$ \cite{kirklin2017eighth}. Consequently, when developing a classifier for mortality, there will be a dearth of training data for the minority class.

 Such imbalanced data is very common in real-world domains like various types of fraud detection \cite{abdallah2016fraud}, and  medical diagnosis of rare diseases  \cite{mazurowski2008training,zhang2018imbalanced,gao2018predicting,fotouhi2019comprehensive}. Accordingly, there has been significant research about techniques to compensate for the data imbalance \cite{fernandez2018learning,lopez2013insight,krawczyk2016learning}. However, when \textit{evaluating} the final classifier on real data, the dilemma of imbalance remains, which can adversely affect the traditional measures of performance, such as the ROC. Consequently ROC can portray an overly-optimistic performance of the model \cite{weng2008new,berrar2012caveats,saito2015precision,davis2006relationship}. This paper illustrates this problem in the context of LVAD classifiers for mortality, and presents and alternative metric that is sensitive to the imbalance in the data, and can better assess the model in predicting the minority class.

%This paper is organized as follows. Section \rom{2} introduces a case study including two classifiers of 90-day mortality. Section \rom{3} describes the issue of imbalanced data.  Section \rom{4} briefly describes the theoretical background of ROC. Section \rom{5} outlines the main limitation of ROC in the case imbalanced data. Section \rom{6} introduces a more effective metric for imbalanced data.  Sections of \rom{7}  and \rom{8} are the discussion and conclusion, respectively.

\section{Methods and Background}
\subsection{Comparison of Two Classifiers for 90-day Mortality}
This study considered the performance of two classifiers for predicting 90-day mortality after LVAD implantation: The HeartMate Risk Score (HMRS) and a Random Forest (RF) that was derived de novo. The HMRS computes the 90-day risk scores (high, medium, and low risks) for mortality based on the five variables including: age, albumin, creatinine, INR, and center LVAD volume\cite{cowger2013predicting}. This score was derived by logistic regression from 1,122 patients who received a HeartMate II as a bridge to transplant or destination therapy \cite{cowger2013predicting}.

The RF is a popular ensemble algorithm constructed by combining multiple decision trees made based on ``bootstrap'' samples from data with random feature selection \cite{breiman2001random}. Each tree in RF will have a vote for a patient, then the output of the classifier is determined by majority voting of the trees. For this study, a RF was derived based on 235 pre-LVAD clinical variables, such as  lab values, demographic information, patients' clinical history, etc., recorded from 11,967 patients with advanced heart failure who received a continuous-flow LVAD recorded in the Interagency Registry for Mechanically Assisted Circulatory Support (INTERMACS). The data were randomly divided into a training ($70\%$) and a test ($30\%$) set. The HMRS score was computed for a subset of the test data set, censoring patients who received a heart transplant or had total recovery before 90 days, and for whom the data records did not contain all five variables required to compute HMRS \cite{cowger2013predicting}. The resulting data set for computing HMRS included 800 patients (mean age of 59 years; 146 females vs. 654 males), extracted from INTERMACS.

%The output of HMRS is a continuous score that is segregated into 3 risk categories: low risk ($<10\%$), medium risk ($10\%$ to $20\%$), and high risk ($>20\%$).\cite{} These categories highly depend on the chosen risk-score thresholds wherein a small change in the thresholds could lead to very different prediction results. To avoid unambiguous assessments, this study examine the power of pure continuous HMRS in predicting the actual observed mortality over all potential threshold. In this way, the HMRS's outcome can be fairly compared against other classifiers such as RF with different range of continuous outcomes. 

The majority of the 800 patients used in this study survived to 90 days (SURV class) $92\%$  and only $8\%$ of patients were dead at 90 days (DEAD class). Thus, there is a high imbalance between the SURV class (majority class) and DEAD class (minority class) in these data.

\subsection{The Problem of Imbalance (and Overlap)}
Without loss of generality, a classifier provides a predicted probability of a class, or label. This is known as \textit{soft classification}, for example the  predicted probability of a hypothetical patient being dead is 30$\%$. Then, this predicted probability can be transitioned into the form of a \textit{hard label}, for example assigning the label ``DEAD'' to the patient. The simple way of bridging from soft classification to hard labels is to set a cutoff or threshold, as shown in Fig 1.a. For example, if the predicted probability of being dead (PPD) is greater (less) than the cutoff point, then the hard label is DEAD (SURV). Then, to evaluate the classifier's performance, the predicted label is compared to the actual outcome and summarized in the form of confusion matrix, as shown in Fig 1.a (inset) containing four elements: True Dead, False Dead, True SURV, False SURV. From these four elements, several evaluation metrics can be computed, including sensitivity, precision, and specificity.

\begin{figure}[b]
	\begin{subfigure}[t]{0.47\textwidth}\centering
		\centering
		\includegraphics[width=0.85\textwidth,height=0.2\textheight]{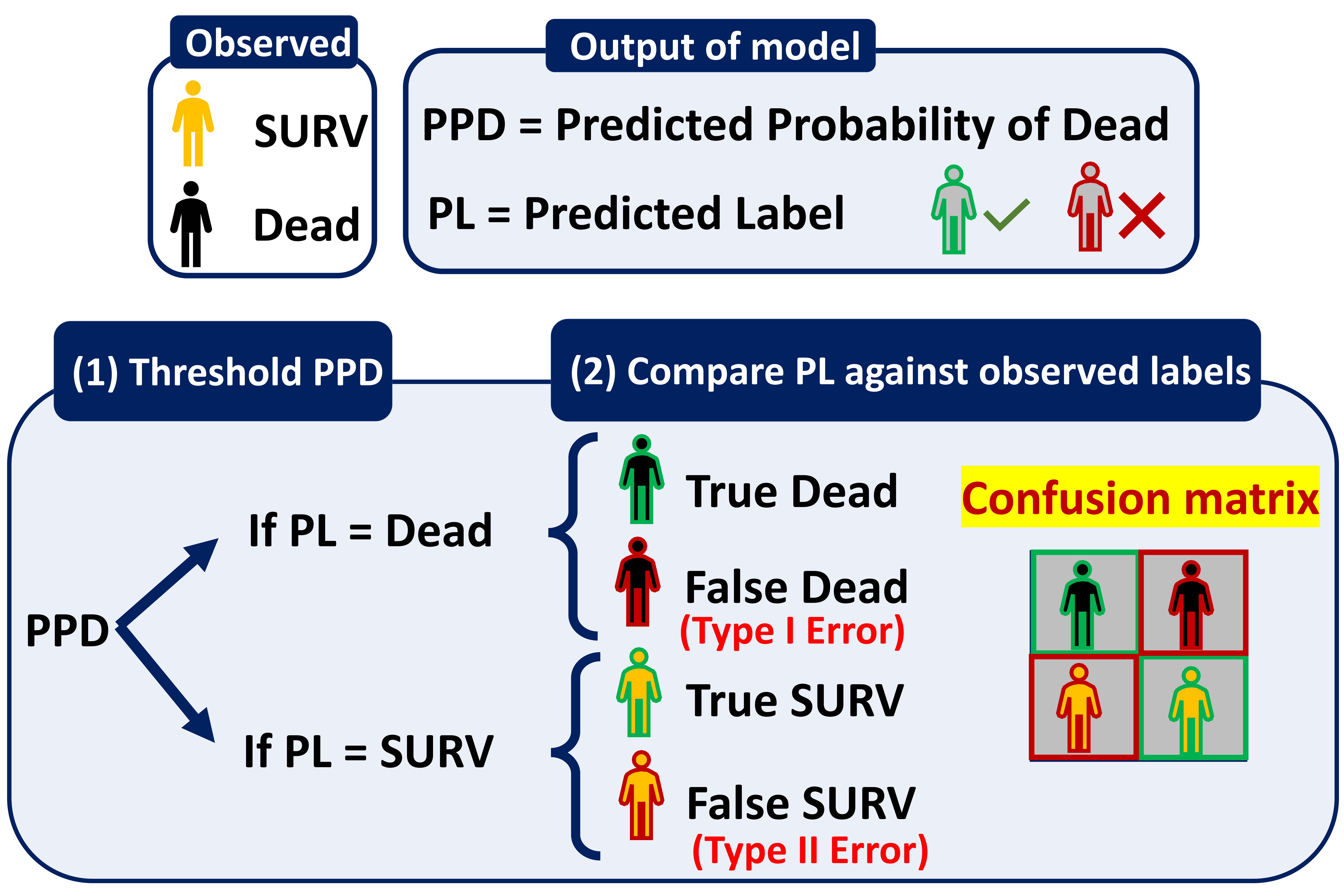}
		\caption{Transition of the outcome of a classifier from Predicted Probabilities of Dead (PPD) to Predicted Label (PL)- (1) Threshold the PPDs: If the PPD for a patient is greater than the threshold then the PL would be Dead otherwise PL would  be SURV. (2) Compare generated PL against the observed class/label and form the confusion matrix.}
	\end{subfigure}
	 \hfill
	\begin{subfigure}[t]{0.47\textwidth}\centering
		\centering
		\includegraphics[width=0.85\textwidth,height=0.2\textheight]{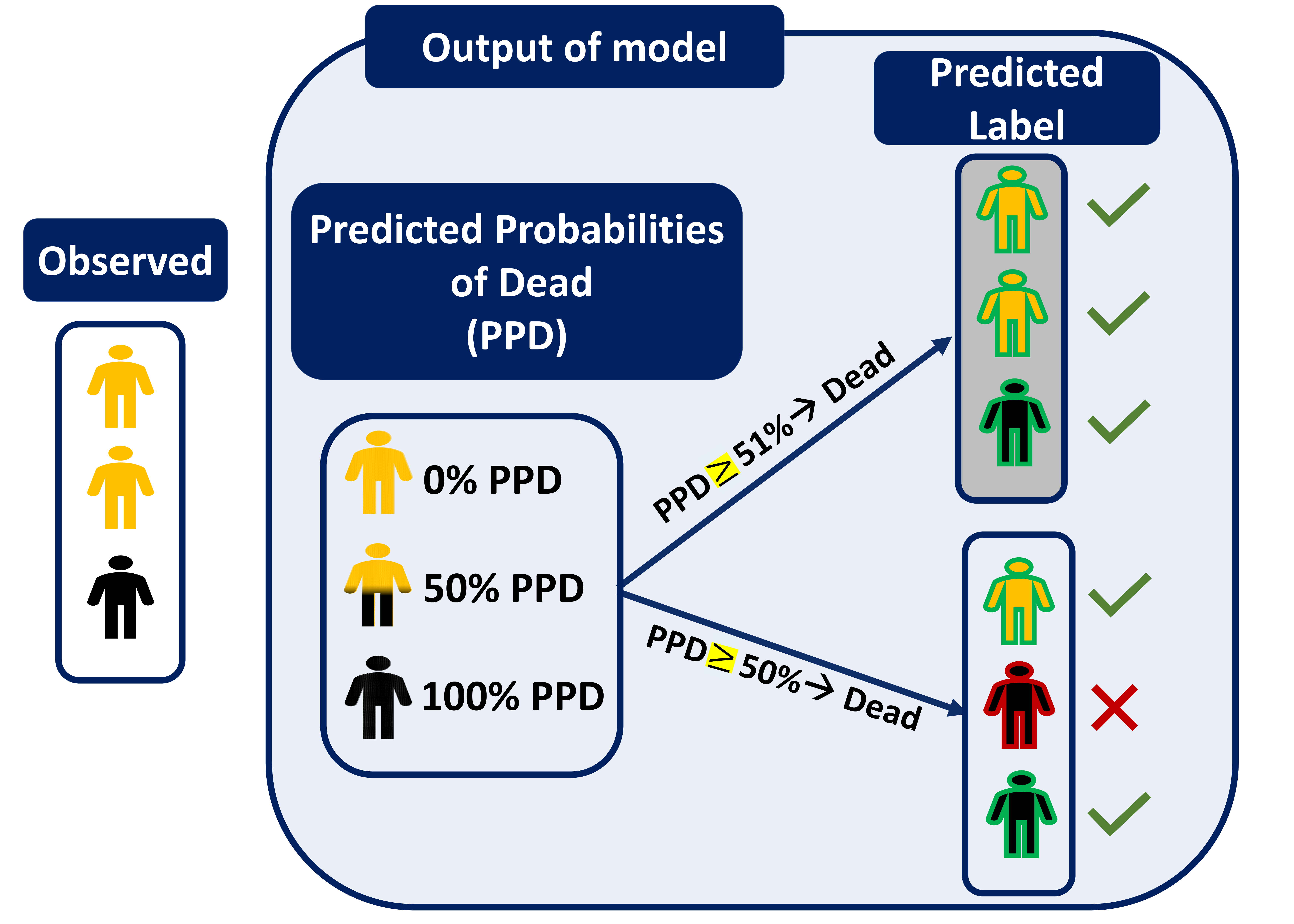}
		\caption{An example shows the transition of the outcome of a classifier for three patients from PPDs to PLs using two slightly different thresholds- The PLs generated using the threshold of $51\%$ (with gray background) are all correctly classified vs the threshold of $50\%$  caused one misclassified label}
	\end{subfigure}
	\caption{}
\end{figure}

Choosing the best threshold is a challenging task that can highly affect the perception of model's performance. For example, Fig 1.b shows the predicted probabilities of being dead for three patients based on the output of a classifier using two potential thresholds of  $50\%$ and $51\%$. In this example the hard labels for the two extreme cases (PPD = $0\%$ and $100\%$) are unambiguous. However, the patient with in the middle with the $50\%$ predicted probability of being dead (hence $50\%$ of chance of being alive) can be classified with either the hard label SURV or DEAD by merely altering the threshold by one percentage point (from $50\%$ vs. $51\%$). In Fig 1.b, to optimize the performance of this classifier, the $51\%$ threshold leads the correct identification of all three patients. However, optimizing the threshold is not as straightforward as this example.

More generally, when considering a larger population of patients, say n=250 (See Fig 2.a, left), the distribution of PPD contains an ambiguous overlap in which an intermediate range of probabilities is associated with both classes. (See Fig 2.a, bottom right). This is in contradistinction to the ``perfect" classifier that does not contain any such overlap. (See Fig 2.a, top right). Therefore, choosing the threshold involves a subjective trade off decision: which is worse, incorrectly predicting a patient as being dead (False DEAD, type \rom{1} error), vs incorrectly predicting a patient as alive (False SUVR, type \rom{2} error)?

\begin{figure}[h]
	\centering
	\begin{subfigure}{0.5\textwidth}\centering
		\centering
		\includegraphics[width=0.95\textwidth,height=0.20\textheight]{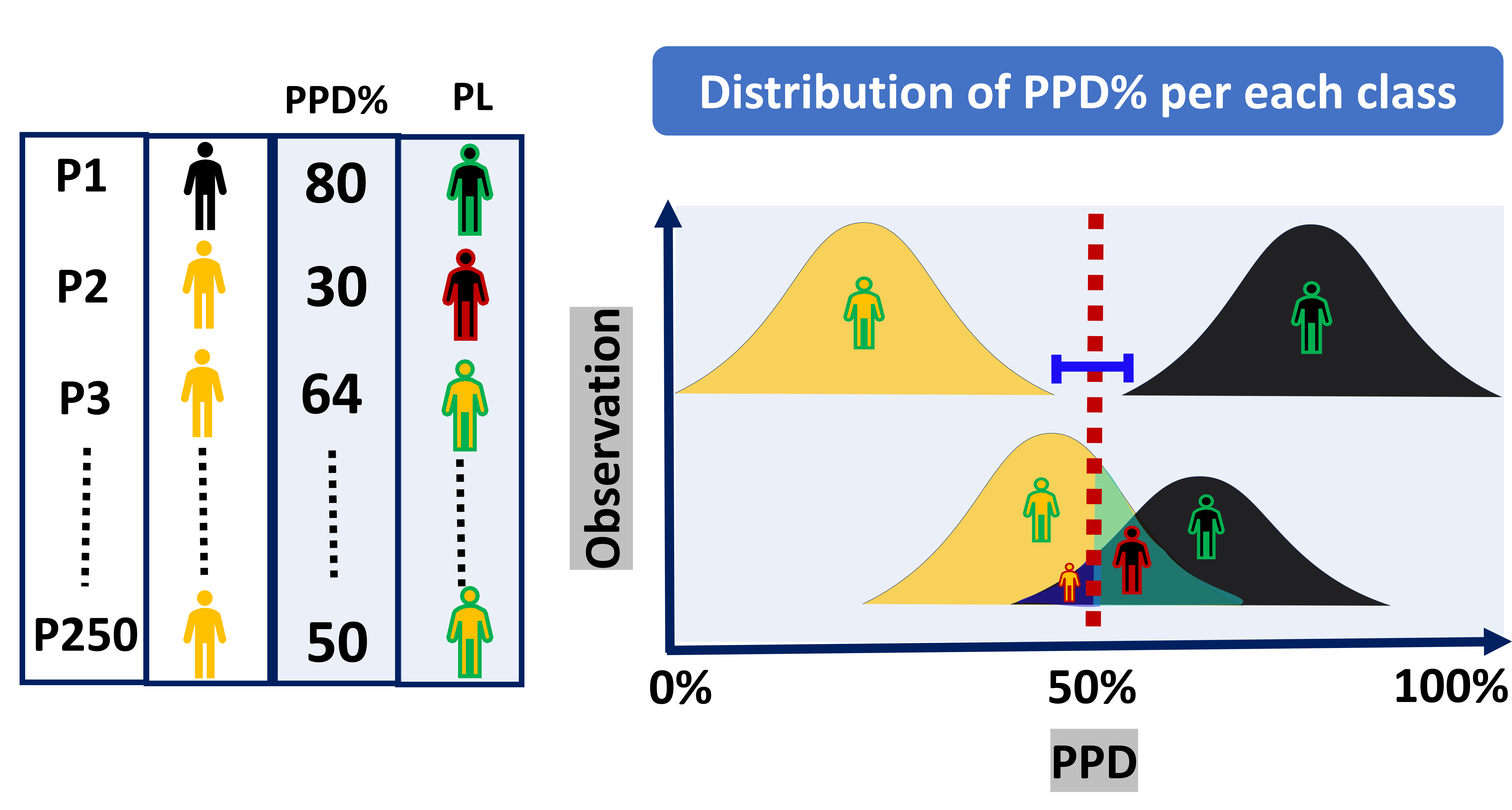}
		\caption{Overlapping issue-Right Figure: A theoretical example of a classifier's output for 250 patients including: Predicted Probabilities of Dead (PPD) and Predicted Labels (PL). Left Figure: Top plot shows the outcome of a perfect classifier with no overlapping between the distributions of PPD of Dead class (colored in black) and SURV class (colored in orange).  Bottom plot shows an imperfect classifier that generates PPDs with ambiguous overlap in which an intermediate range of probabilities is associated with both (either) class.\\}
	\end{subfigure}\\
	\begin{subfigure}{1\textwidth}\centering
		\centering
		\includegraphics[width=0.85\textwidth,height=0.25\textheight]{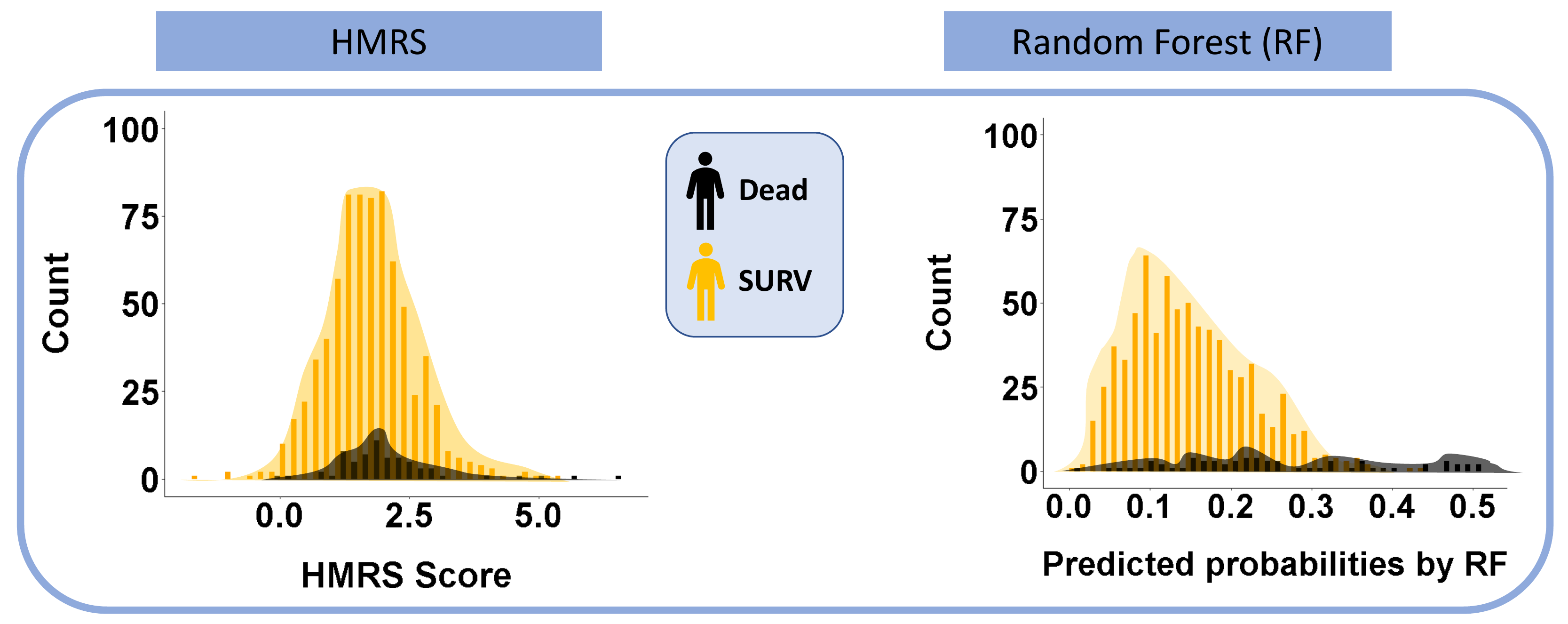}
		\caption{Overlapping issue for the outcomes of 90-Day LVAD mortality classifiers- The underlying distributions of HMRS  risk scores and  RF predicted probabilities of DEAD for 800 patients in this study data are shown histogram plots. The histograms are categorized based on the observed labels for patients in this study.}
	\end{subfigure}
	\caption{}
\end{figure}	

When the data are highly imbalanced, the unequal distribution of classes will compound the problem of overlap and make classification an even more challenging task. The LVAD 90-day mortality study introduced in the previous section is an example. Fig 2.b shows the histograms of HMRS Risk score (left plot) and RF probability of mortality (right plot) categorized by their actual mortality outcome (DEAD vs SURV). The predicted probabilities of being dead generated by RF for all 800 patients in this study (total of both DEAD and SURV) ranges from 0.01 to 0.52 with the mean of 0.15; while HMRS scores range between -1.80 to 6.50 with the mean of 1.71. The means of both distributions are closer to the lower part of their ranges because of the preponderance of alive patients ($92\%$) in the test sample. This is clearly visible in Fig 2.b. as the black bars (DEAD class) are much lower than the orange bars (SURV class.) HMRS has a tall bell shape distribution of scores around the means for both classes. On the other hand, the distribution of the SURV class in the RF histogram is right skewed, whereas the distribution of the DEAD class is relatively flat, with not identifiable maximum. However, RF was more successful in separating the classes for two reasons: first, the SURV class is clustered within a narrow band (approximately 0.0 to 0.3; median of approximately $10\%$.) Secondly, the band above  $35\%$ is dominated by the Dead class. Nevertheless, for both of the plots in Fig 2.b, optimizing a cutoff point threshed that efficiently separate both classes is a challenging task. By default, any cutoff point threshed for both plots in Fig 2.b will result in a much greater number of True SURV and False SURV comparing to True DEAD and False DEAD. On account of this dilemma (imbalance plus overlap), caution is needed when applying metrics of performance to these classifiers - such as the common ROC.

\subsection{Receiver  Operating  Characteristic (ROC)}
The ROC curve is defined as the LOCUS of  True Positive Rate (\textbf{\textit{TPR}}) and False Positive Rate (\textbf{\textit{FPR}}) for\textbf{\textit{ all possible choices of cutoff thresholds}} as shown in Fig 3.a. The color bar to the right of the ROC curve represents the threshold levels. Another term for TPR is sensitivity or recall. Another term for FPR is 1-Specificity or 1-True Negative Rate (TNR).  The Y-axis (TPR) is the proportion of True Dead over all observed Dead.  The x-axis (FPR) is one minus the proportion of True SURV over the total of observed SURV. In other words, ROC looks at the performance of a classifier in prediction of both of classes, Dead and SURV.  The \textit{overall} performance, for \textit{all} threshold values, can be assessed by computing the Area Under the Curve of ROC (AUC-ROC).

The shape of the ROC curve depends on the overlap between the distributions of predicted probabilities of the two classes. A perfect classifier with no overlapping  will have an L-shape ROC curve (dashed line in Fig 3.a)  with AUC-ROC equal to one, and will pass through the point of (FPR=0, TPR=1), as indicated with the gray dot, corresponding to the threshold where all patients are correctly identified by the classifier. The closer the ROC curve gets to the diagonal ``line of unity'' (representing random chance) such as the solid curve in Fig 3.a the worse the performance of classifier.  The purple and red dots correspond to the upper and lower bounds of the threshold: $>1$ and $0$, respectively. At the lower bound (threshold = $0$), the  classifier identifies all patients as SURV (FPR= $0\%$ or specificity=$100\%$). At the upper bound, the classifier identifies all patients as DEAD (sensitivity= $100\%$).

%In this example, there are three colors in the ROC curve, which is equal to the number of unique values of PPD for three patients in Fig.2. Each color corresponds to one cutoff point per patient that changes the predicted label for that patient, and thus changes the TPR and FPR of the ROC curve. For instance, the green color represents the threshold of $50\%$ for the alive patient with PPD of $50\%$ in Fig.2 which leads this patient to be incorrectly labeled as Dead resulting in an increase of FPR  from $0\%$ to $50\%$  while sensitivity remains $100\%$ as the classifier still correctly identifies the only dead patient as DEAD. 

%This is because the ROC curve in Fig.5 is plotting the performance of the classifier in Fig.2 only for three patients with Predicted Probabilities of Dead (PPD) of $0\%$, $50\%$, and $100\%$ that changing threshold between all the possible cutoff points between $0\%$ and $100\%$ would not change the value of specificity and sensitivity, and thus not changing the X or Y values on ROC curve.

%Both of sensitivity and specificity would be one at $50\%$ (presented by a cyan dot on ROC in Fig.5) but for a range of thresholds less than $50\%$ (presented by green color on ROC in Fig.5), such as cutoff point of $49\%$ that was shown previously in Fig.2, the alive patient with PPD of $50\%$ will be incorrectly labeled as Dead resulting in specificity reduced to $50\%$ (FPR = $50\%$) but sensitivity will remain $100\%$ as classifier still correctly identified the only dead patient as Dead. 

\begin{figure}[h]
	\begin{subfigure}{0.5\textwidth}\centering
		\centering
		\includegraphics[width=1\textwidth,height=0.4\textheight]{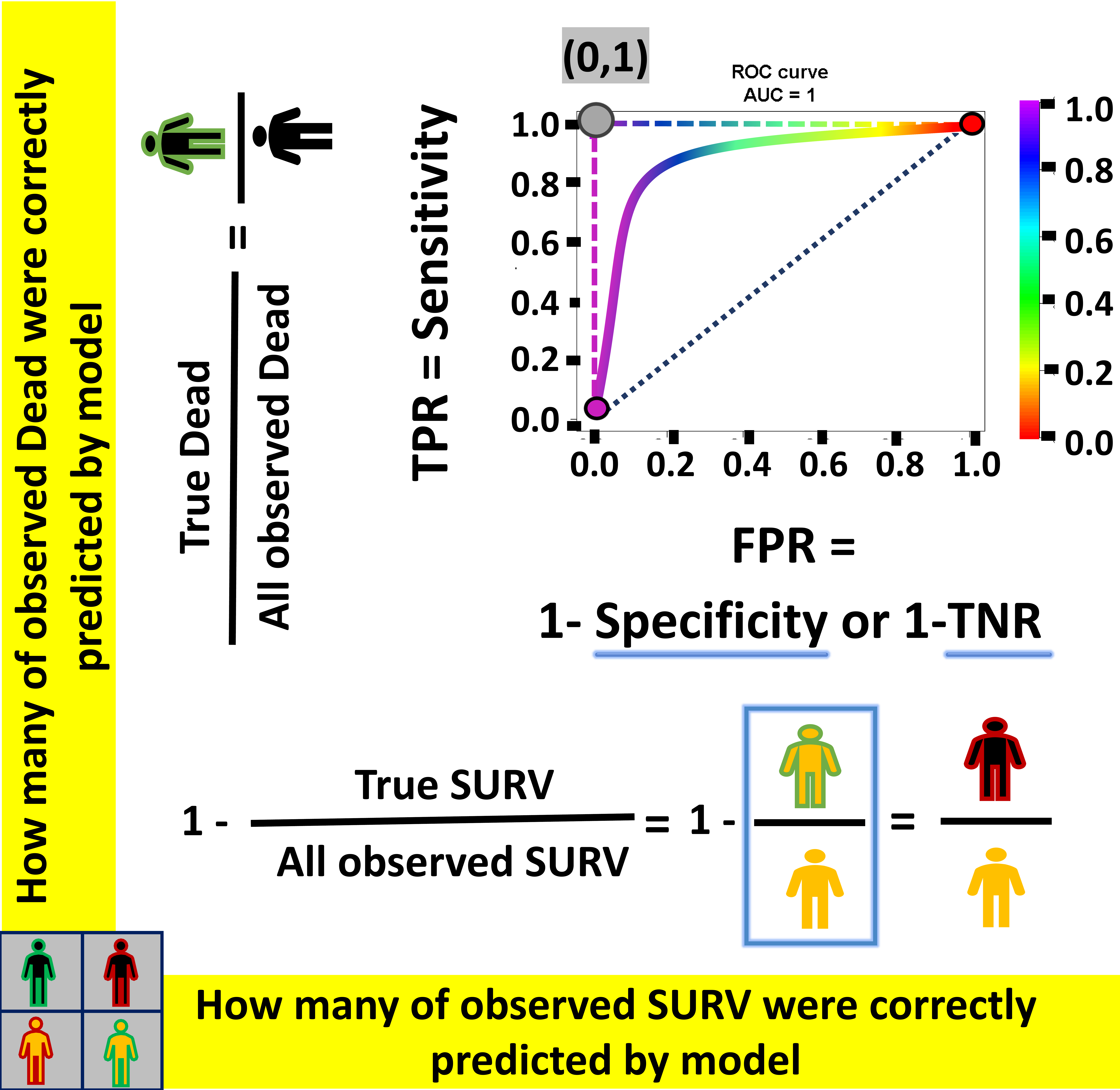}
		\caption{ROC- The example of ROC curves for a perfect classifier (L-shape dashed-curve), an imperfect classifier (solid curve), and a random classifier (diagonal dotted-line).}
	\end{subfigure}\hspace{7mm}
	\begin{subfigure}{0.5\textwidth}\centering
		\centering
		\includegraphics[width=1\textwidth,height=0.4\textheight]{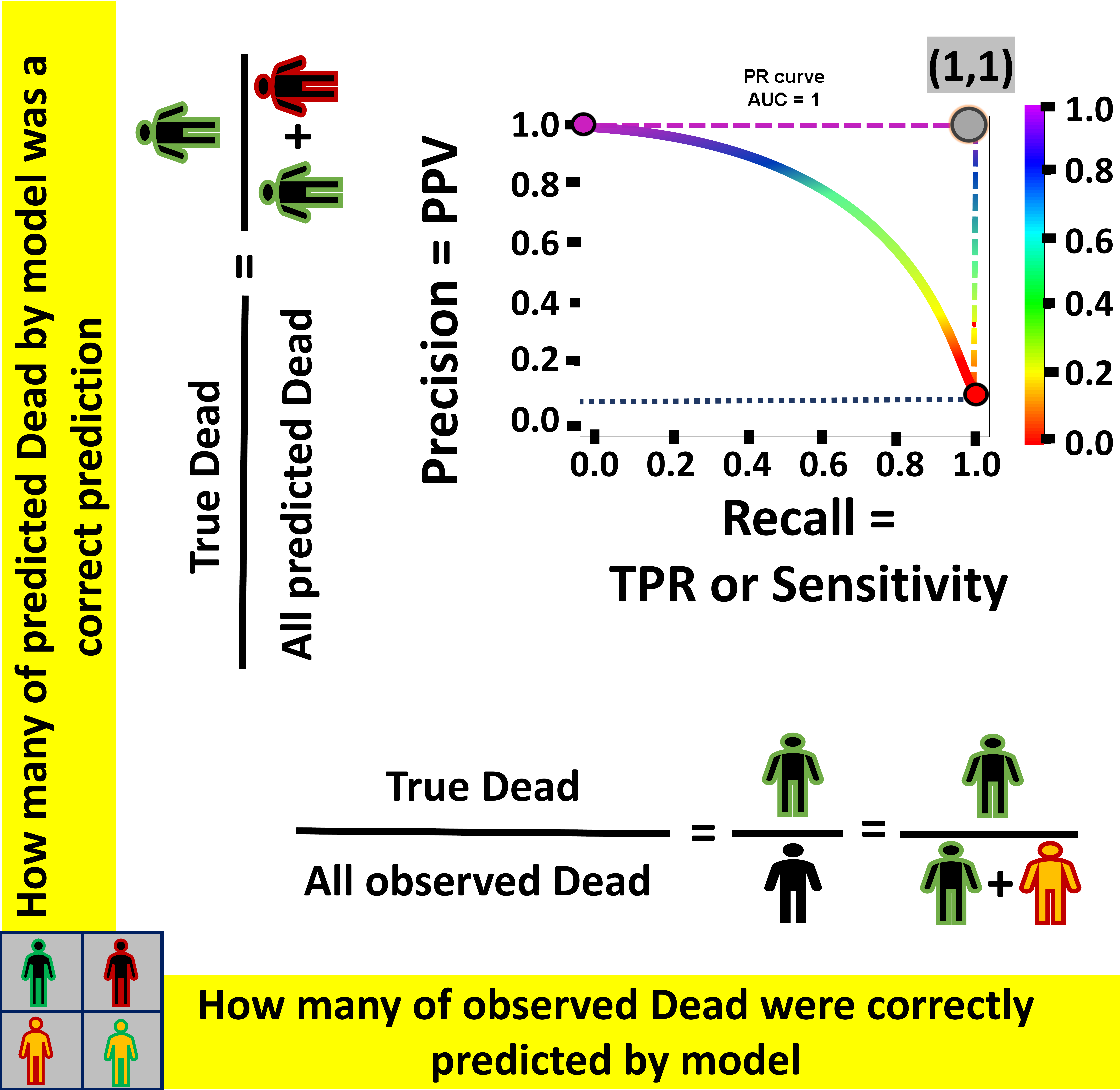}
		\caption{PRC-The example of PRC curves for a perfect classifier (L-shape dashed-curve), an imperfect classifier (solid curve), and a random classifier (diagonal dotted-line).}
	\end{subfigure}
	\caption{}
\end{figure}

\subsection{Precision-Recall Curve (PRC)}

The PRC is a plot the LOCUS of  \textbf{\textit{Precision}} and \textbf{\textit{Recall}} for \textbf{\textit{all possible choices of cutoff threshold}} as pictured in Fig 3.b. The X-axis in the PRC (recall) is the same as the Y-axis in the ROC. Other terms for recall are sensitivity and TPR, equal to True DEAD over all observed Dead. The Y-axis (precision) is also known as positive predictive value (PPV), equal to True DEAD over all predicted DEAD by the model. The color bar to the right of the PRC curve represents the threshold levels of the classifier. It is important to note that the formulas for  precision and recall have the same numerator (True DEAD) and both include True DEAD in their denominator. Their only difference is one element in their denominators: False DEAD in precision and False SURV in recall. Thus, PRC focuses on the quantity of True DEAD over various cutoff thresholds considering errors of both classes: False DEAD and False SURV. Therefore, PRC is beneficial when dealing with imbalanced data as it focuses on the performance of model in only the minority class (DEAD) and it is sensitive to skewness in the imbalance data. Similar to ROC, the PRC can be summarized by computing the Area Under Curve of PRC (AUC-PRC). In PRC, the random classifier would be a horizontal line with precision equal to proportion of minority class; for instance, $8\%$ for 90-day post-LVAD mortality.

A perfect classifier  will have an L-shape PRC curve (dashed line in Fig 3.b) with the AUC-PRC of 1, and will include the point (recall= 1, precision= 1), as indicated with the gray dot in the figure,  corresponding to the cutoff threshold where all patients are correctly identified by the classifier. However, when there is overlap between predicted probabilities of classes, as illustrated in Fig 2.b, the  PRC curve approaches the dotted horizontal line corresponding to a random classifier. The red and purple dots in Fig 3.b corresponds to the two extreme thresholds. The red dot (threshold = 0.0) indicates the point where recall = 0, and precision =0/0, indicated by 1.0. The purple dot corresponds to the threshold 1.0 wherein the classifier identified all patients as DEAD; hence the recall $=$ 1.0 (no False SURV) and the precision $=$ overall proportion of the minority (DEAD) class.

%The start point of PRC is (recall= 0, precision= 0/0) where the classifier identifies all patients as SURV, resulting in no False DEAD and no True DEAD. The end point of PRC is (recall= 1, precision= proportion of minority class-DEAD class) where classifier identifies all  patients as DEAD resulting in no False SURV.

%with precision equal to $33\%$
%The  AUC-PRC varies  with class skew in imbalanced data \cite{boyd2012unachievable}.

\section{Result}
\subsection{Limitations of ROC for imbalanced LVAD mortality study}
\begin{figure}[]
	\centering
	\begin{subfigure}{1\textwidth}\centering
		\centering
		\includegraphics[width=0.95\textwidth,height=0.3\textheight]{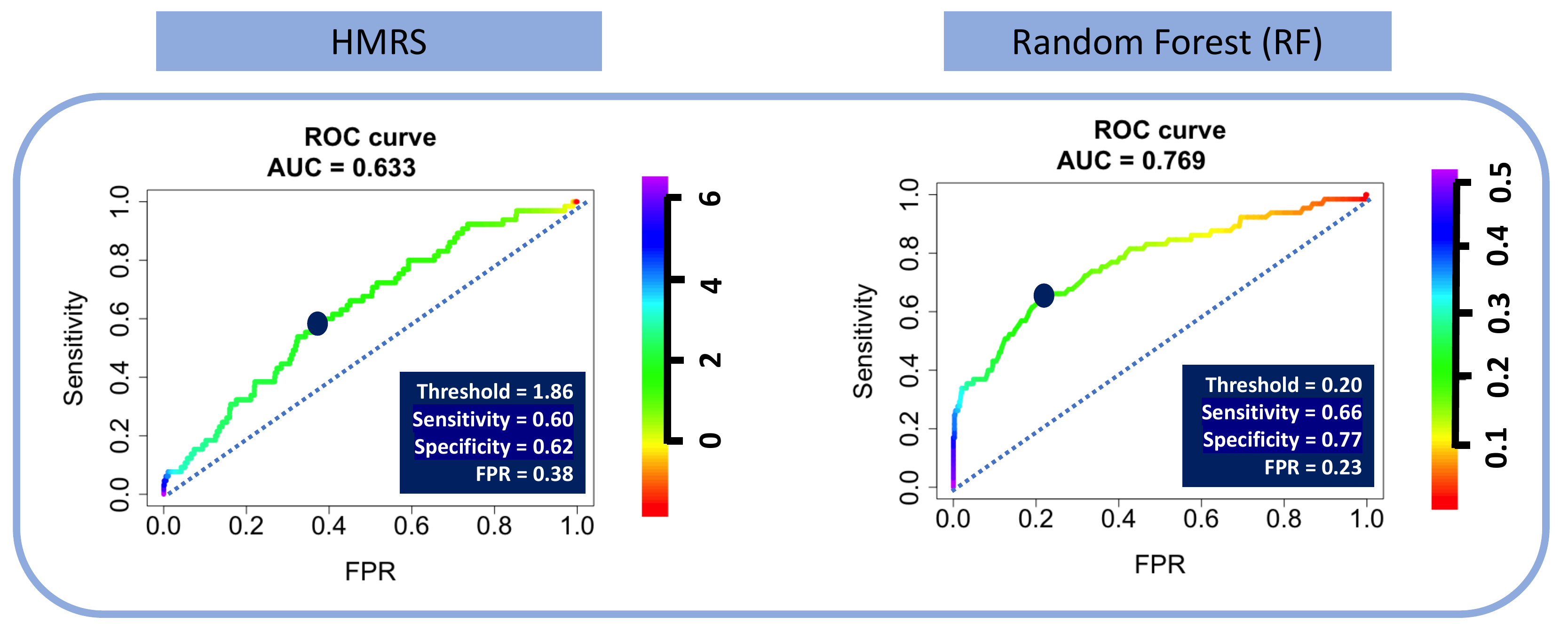}
		\caption{The ROC for HMRS and RF- The dark blue points indicate the optimal cutoff thresholds, detailed in the inset tables.}
	\end{subfigure}
	\par\bigskip
	\begin{subfigure}{1\textwidth}\centering
		\centering
		\includegraphics[width=0.95\textwidth,height=0.4\textheight]{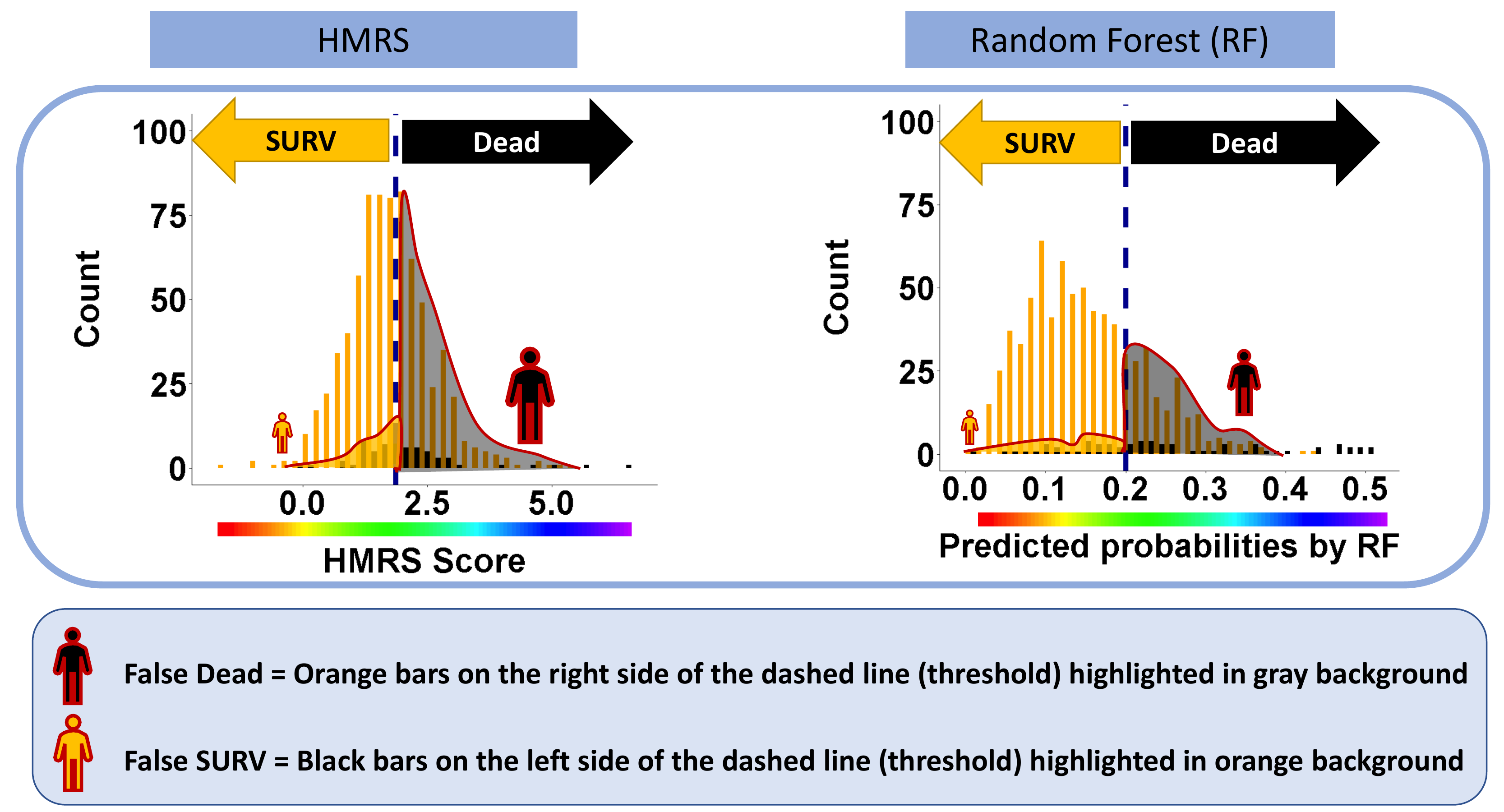}
		\caption{The distributions of false predictions for HMRS  and  RF classifiers- These histograms are the same histograms in Fig 2.b. The dashed lines are corresponding to optimized cutoff thresholds chosen based on ROC curves in Fig 4.a. The two types of errors,  False Dead and False SURV associated with this threshold are reflected in black and orange regions with red outline, respectively.}
	\end{subfigure}
	\caption{}
\end{figure}

Fig 4.a shows the ROC curves for the two classifiers, HMRS and RF, for prediction of 90-day mortality after LVAD implantation. The color of the curves corresponds to the values of cutoff thresholds in the range of each classifier's outcome as presented in their corresponding legends. The  predicted probabilities of being dead by RF for 800 patients in this study  ranges from 0.01 to 0.52 vs HMRS ranges between -1.8 to 6.50. The dominant color in the ROC curve for HMRS is green corresponding to the compact (tall and narrow) distribution of scores around the mean of 1.71 as shown in Fig 2.b.  Therefore, a small change in cutoff threshold around the mean  may  dramatically change the performance of classifier due to high number of patients located around the mean. On the other hand, the ROC curve for RF illustrates the performance of RF over a more uniformly distributed range of thresholds, especially  for the lower part of the range (less than $30\%$), corresponding to the right-skewed distribution of predicted probabilities shown in RF's histogram for SURV class (orange bars) in Fig 2.b. The primary difference between the two ROC curves in this example is that AUC-ROC for RF is slightly greater than the AUC-ROC for HMRS, 0.77 vs. 0.63, indicating better overall performance of RF in separating DEAD vs. SURV. 

The two dark blue points on the curves indicates the optimized threshold points of 1.86 and 0.21 for HMRS and RF, respectively, where the values of sensitivity and specificity are effectively equalized. Although the values of sensitivity for HMRS and RF at the optimized threshold are similar, 0.60 and 0.66, respectively, the corresponding specificity of RF, 0.77, is notably greater than for HRMS, 0.62. Translating these optimized thresholds to histograms of Fig 2.b illustrates the efficacy of each classifier in separating classes. (See Fig 4.b.) Comparison of the two types of errors: False SURV (dead patients incorrectly classified as alive) and  False DEAD (alive patients incorrectly classified as dead) reveals that the proportion of False DEAD is much greater than the proportion of False SURV for both classifiers. This is due to a combination of the imbalance of the data (about $92\%$ alive patients) and relatively poor performance of the classifiers. However, the False DEAD is visibly larger for HMRS  compared to RF due to the huge overlap between  distributions of HMRS scores for DEAD and SURV classes. 

The stark differences revealed by the histograms in Fig 4.b are not discernible from comparison of the corresponding ROC curves. For example, a small change of the  threshold in the ROC curves in Fig 4.a  corresponds to a small change in both Sensitivity and FPR. However, Fig 4.b reveals that the shifting the cutoff from the optimal point (left or right) will result in a much greater change in False DEAD vs False SURV.  This is because the denominator of FPR in the ROC curve plots, the \textit{total} number of SURV which is a huge number, thus attenuating the effect of changes in the numerator, False DEAD. In  terms of the confusion matrix, this can be restated as  number of False DEAD is overwhelmed by the much larger number of True SURV -- considering that the total observed SURV is the sum of True SURV and False DEAD. Consequently, if the performance of RF and HMRS compared by only considering their ROC curves in Fig 4.a, there might not be a huge difference between the performance of two models while  Fig 4.b obviously shows that RF suffers much less from error of False DEAD than HMRS. In addition, it was shown that choosing threshold only based on the ROC curve may cause unintentional effects in the perception of model with respect to the minority class. In conclusion, when the ROC is dominated by the majority class (the large proportion of patients that survive), it poorly reflects the performance of the model with respect to the minority class (dead patients), and thus may be a deceptively optimistic evaluation tool in the case of imbalanced data.

In light of the aforementioned limitation of ROC, there is clearly a need for a \textit{supplemental} evaluation tool that is sensitive to skewness in the data and emphasizes the performance on the minority class. One such evaluation tool is the Precision-Recall Curve (PRC) \cite{saito2015precision,davis2006relationship,cook2020consult}.

%Another comparison can be observed by focusing on the \textit{early retrieval area} of the ROC plot, highlighted in the red shaded region on ROC curves in Fig.6. The early retrieval area in the RF ROC indicates that increasing sensitivity from 0$\%$ to 30$\%$  does not adversely affect the FPR (1-specificity), which remains very close to 0$\%$. This is contrasted to the small early retrieval area of the HMRS.

\subsection{Solution: PRC for imbalanced LVAD mortality study}
 Fig 5 shows the PRC curves for the two classifiers of HMRS and RF for prediction of 90-day mortality after LVAD implant. The color legends indicate the same thresholds values as presented in the ROC curves above (Fig 4.a). Unlike the ROC curve, the distribution of colors is more uniform, visible by the broader spectrum of colors in in the curves, especially the blue colors toward the upper range, which are virtually absent in the corresponding ROC curves in Fig 4.a. This is important because the blue portion of the curve relates to the upper bound of threshold values (high predicted probabilities of being DEAD) in which the classifiers have the greatest precision, i.e. when more of the predicted DEAD by the classifier are True DEAD. This reveals a striking difference between HMRS and RF inasmuch as the blue region (high precision) of HMRS is limited within a very narrow band of recall, and virtually vertical. The PRC reveals that the precision  drops precipitously to approximately $10\%$ (close to the random classifier: blue dotted line) as recall (sensitivity) increase from $0\%$ to about $10\%$. This corresponds to the severe overlap between the classes in the histogram of HMRS (Fig 2.b), even for the highest scores which leads to the huge proportion of False DEAD (alive patients incorrectly identified as DEAD). This is contrasted with the PRC of RF which decreases more gradually in precision with increasing recall. Also, it is noted that the PRC of RF remains at nearly 1.0 over a wider range of threshold, i.e. between $0.44$  ($44\%$) and $0.51$ ($51\%$) corresponding to a range of recall (sensitivity) from $0\%$ to $17\%$. Overall, from the perspective of precision-recall, RF outperforms HMRS with AUC-PRC of 0.43 vs. 0.16.

The dark blue dots in Fig 5 correspond to optimized thresholds chosen based on the ROC curves in Fig 4.a. It is readily seen that the precision of both classifiers at these thresholds is very low, although RF has a better precision, $38\%$, for achieving the sensitivity of $66\%$ than HMRS with precision less than $10\%$ for sensitivity of $60\%$. Using these thresholds, the HMRS classifier will correctly identify only 38 out of 64 dead patients ($60\%$ sensitivity) in the 800-patient test data set, yet will \textit{incorrectly} label 308 patients ($90\%$ of the 342 patients labeled as DEAD) that are actually alive!

On the other hand, if we assert that precision and recall are equally important, the corresponding optimal cutoff would be indicated by the red dots on PRC curves in Fig 5. for which both precision and recall of HMRS and RF is $15\%$ and approximately $38\%$, respectively. At these optimized points, the  harmonic mean of precision and recall (F1-Score) equals to both precision and recall. These optimized points are not necessary the best way of choosing the threshold since it results in very low levels of recall. However, it illustrates that the choice of threshold highly depends the comparative ``importance'' of sensitivity and precision; hence acceptance/consequence of errors (False DEAD vs False SURV).

\begin{figure*}[h]
	\centering
	\includegraphics[width=1\textwidth,height=0.3\textheight]{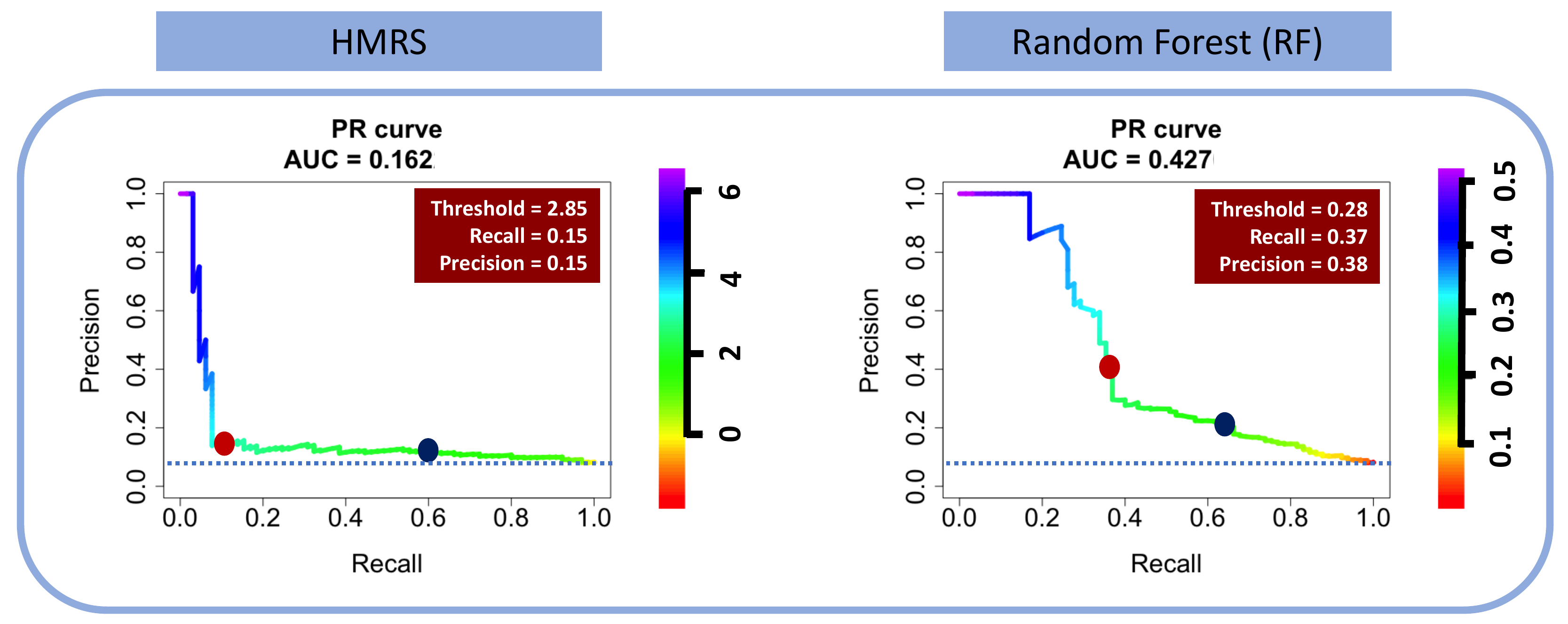}
	\caption{The PRC for HMRS and RF classifiers- The dark blue point on the PRC curves are corresponding to optimized thresholds chosen based on the ROC curves in Fig 4.a. The red boxes are the corresponding specifications of  red dot points on PRC curves presenting the optimized cutoff thresholds of PRC curves.}
\end{figure*}

\section{Discussion}
The clinical utility of a risk score or classifier for mortality following LVAD implantation depends greatly on the degree of separability between predicted probabilities of the two classes: DEAD vs SURV. (See Fig 2.a). Overlap between the distributions of classes creates an intermediate range of probabilities that is associated to both classes. This results in two types of errors: False DEAD (alive patients who are incorrectly labeled as DEAD) and False SURV (dead patients who are incorrectly labeled as SURV). Therefore, the choice of a threshold is tantamount to choosing between these two types of error. This dilemma is accentuated when the data are highly imbalanced, as is the case of 90-day mortality post-LVAD. (See Fig 2.b). The overwhelmingly large size of the majority class, SURV class, amplifies the False DEAD error much more than False SURV. (See Fig 4.b). Thus, when choosing a threshold and evaluating the performance of these classifiers, it is very important to focus on the minority class (both True DEAD and False DEAD). 

This study illustrated that the ROC, a well-known evaluation tool used for most LVAD risk scores, in the case of imbalanced data leads to an optimistic perception of the performance of the classifier. This is due to the intuitive but misleading interpretation of specificity: where the large number of False DEAD error is overwhelmed by the huge number of All observed SURV in its denominator. Neglecting the full magnitude of False DEAD generated by a classifier or risk model, i.e \textbf{\textit{precision}}, could give the clinician false confidence in the prediction of DEAD by the classifier. Unfortunately, most of published pre-LVAD risk scores and classifiers have not reported their precision. Therefore, these scores should be used with extreme caution.

The \textbf{\textit{Precision Recall Curve (PRC)}} was shown here to be a useful tool to evaluate and compare the performance of two different risk classifiers for 90-day mortality following LVAD implantation: The HeartMate Risk Score (HMRS) and a 
random forest (RF) classifier derived from INTERMACS, a highly imbalanced data set. The PRC plots the proportion of True DEAD to both of the errors: False DEAD and False SURV. Thus, PRC emphasizes the performance of a classifier for the minority class (DEAD class), in contradiction with the ROC which has an equal emphasize on both minority and majority classes. PRC is not affected by the overwhelming number of True SURV (majority class), and thus it does not generate a misleadingly optimistic perception performance, as does the ROC. 

The preceding is not an indictment of ROC, but a revelation that ROC fails to paint a complete picture of a classifiers' performance. Therefore, ROC provides a view of classifiers' performance with both minority and majority classes while PRC provides a view of classifiers' performance on minority class which becomes more important and informative when dealing with imbalanced data.

\subsection*{Limitations}
The problem of classifier development with imbalanced data is well-known area of research in many disciplines, including medicine \cite{mazurowski2008training,zhang2018imbalanced,gao2018predicting,fotouhi2019comprehensive}. Accordingly, there exists a variety of approaches to mitigate the effects of imbalance such as resampling methods, assigning weights to minority samples, one-class classifier, etc. \cite{fernandez2018learning,lopez2013insight,krawczyk2016learning}. This study did not attempt to employ any of these methods; however, it would be beneficial in future studies to explore various strategies to achieve the best performance of LVAD classifiers.

\section{Conclusion}
ROC has become a common evaluation tool for assessing the performance of classifiers and risk scores for LVAD mortality. However, because there is a great paucity of dead patients versus alive patients, ROC can provide a misleading optimistic view of the performance of the classifiers in predicting death versus survival. PRC is suggested as a supplementary evaluation tool to address the imbalance in these data.

\section{\textbf{Clinical perspectives}}
Using any classifier for mortality following LVAD implantation inevitably involves choosing a threshold. From a clinical perspective this translates to a conscious decision between risk of inserting an LVAD in a patient who will die due to misplaced faith in the classifier (False SURV); versus denying a patient from a potentially life-saving LVAD because of a false presumption of death (False DEAD) by the classifier. This is not necessarily an easy decision. If the clinician chooses a conservative threshold, so as to avoid False SURV, he/she will mitigate the risk of accelerating a patient's death by inserting and LVAD, however he/she is at a loss for a classifier to evaluate the alternatives. This situation begs for a more holistic approach to stratification of patients with severe heart failure, to provide comparison, or ranking of alternatives, for example the use of a temporary support device as a bridge to VAD. 
%consider adding sentence that alludes to the trust issue; or comprehensibility issue. Risk score makes logical sense to the clinician; the massive black box models are not easily trusted. (maybe move to discussion).

Another consideration that highly affects the tradeoff between False SURV (False Negative) and False DEAD (False Positive) is the intended role of classifier in the clinical assessment of the pre-LVAD patients.  For example, the initial screening test for HIV has a high sensitivity because of the importance of avoiding False Negative. But among those with positive initial screening test, there exists patients who do not actually have HIV (False Positive). Thus, patients with positive initial tests are reassessed with highly precise diagnostic test with low False Positive rate to confirm the HIV diagnosis. Therefore, as a \textbf{\textit{screening tool}}, sensitivity is most important (avoiding False Negative); but as a \textbf{\textit{diagnostic tool}}, precision is more important, to avoid False Positives. By analogy to the pre-LVAD classifier, the choice of threshold might be situation-specific: more conservative as a screening tool, and less so as a definitive diagnostic tool.  In conclusion, there is a need for future studies to comprehensively investigate the role of pre-LVAD risk assessment in clinical medical decisions by considering all-inclusive aspects of clinical settings of pre-LVAD.

\section{Acknowledgment} 
 This work was supported by the National Institutes of Health under Grants R01HL122639 and R01HL134673. Data for this study were provided by the International Registry for Mechanical Circulatory Support (INTERMACS), funded by the National Heart, Lung and Blood Institute, National Institutes of Health and The Society of Thoracic Surgeons.

\bibliographystyle{IEEEtran}
\bibliography{libraryabv}
\vspace{1em}

\end{document}